\documentclass[usenatbib, useAMS, letter]{mnras}

\usepackage{graphicx}
\usepackage{amsmath,paralist,xcolor,xspace,amssymb}
\usepackage{times}
\usepackage{enumitem}

\newcommand{\msun}{{\rm{M}_\odot}}

\newcommand{\lcdm}{$\Lambda$CDM\xspace}

\newcommand{\eagle}{\textsc{eagle}\xspace}

\def\gs{\mathrel{\raise1.16pt\hbox{$>$}\kern-7.0pt \lower3.06pt\hbox{{$\scriptstyle \sim$}}}}         
\def\ls{\mathrel{\raise1.16pt\hbox{$<$}\kern-7.0pt \lower3.06pt\hbox{{$\scriptstyle \sim$}}}}         

\title[The offsets between galaxies and their dark matter]{The offsets between galaxies and their dark matter in \lcdm}

\author[M. Schaller et al.]  {Matthieu Schaller\thanks{E-mail: 
matthieu.schaller@durham.ac.uk}, Andrew Robertson, Richard Massey,
Richard G. Bower \& \newauthor Vincent R. Eke \\
Institute for Computational Cosmology, Durham University, South Road, 
Durham DH1 3LE , UK}
\begin{document}

\date{\today}

\pagerange{\pageref{firstpage}--\pageref{lastpage}} \pubyear{2015}

\maketitle

\label{firstpage}

\begin{abstract}
We use the ``Evolution and Assembly of GaLaxies and their Environments'' (\eagle)
suite of hydrodynamical cosmological simulations to measure offsets between the
centres of stellar and dark matter components of galaxies. We find that the vast
majority ($>\!95\%$) of the simulated galaxies display an offset smaller than the
gravitational softening length of the simulations (Plummer-equivalent
$\epsilon\!=\!700$\,pc), both for field galaxies and satellites in clusters and
groups.  We also find no systematic trailing or leading of the dark matter along a
galaxy's direction of motion.  The offsets are consistent with being randomly drawn
from a Maxwellian distribution with $\sigma\le\!196$\,pc.  Since astrophysical
effects produce no feasible analogues for the $1.62^{+0.47}_{-0.49}$\,kpc offset
recently observed in Abell~3827, the observational result is in tension with the
collisionless cold dark matter model assumed in our simulations.
\end{abstract}


\section{Introduction}
\label{sec:introduction}
Observations of one galaxy in cluster Abell\,3827 (redshift
$z\!\approx\!0.1$; \citealt{2010ApJ...715L.160C}) revealed a surprising
$1.62^{+0.47}_{-0.49}$\,kpc (68\%\,CL) offset between its dark matter and
stars \citep{Massey2015}.  
Such offsets are not observed in isolated field
galaxies \citep{2006ApJ...649..599K,2007ApJ...667..176G}\footnote{ A small number of
galaxy quad lenses are not well-fitted by standard parametric models of dark matter
centred on the optical emission.  To fit lens RXS J1131, \cite{2006A&A...451..865C}
need to include a $0.044\arcsec$ offset or $m\!=\!4$ octupole term.  With lens COSMOS
J09593, \cite{2008MNRAS.389.1311J} achieved an acceptable goodness of fit only with a
$0.063\arcsec$ offset and (an unrealistically large) external shear $|\gamma|=0.25$.
However, in these isolated lenses the cause of these poor fits is more likely to be
local substructure \citep{2013ApJ...767....9H}.  An offset between mass and light
would produce a relatively shallow core profile and possibly more detectable central
images.  Note also that the location of mass peaks is determined much more precisely
by strong lensing than by weak
lensing \citep{2012ApJ...757....2G,2012MNRAS.419.3547D}.}.  However,
offsets inside clusters are consistent with theoretical predictions from models of
self-interacting dark matter \citep[SIDM][]{2000PhRvL..84.3760S}.  As galaxies move
through a cluster core, interactions with the cluster's dark matter would create a
friction and cause a galaxy's dark matter to lag slightly behind its
stars \citep{2011MNRAS.413.1709M, 2014MNRAS.437.2865K, 2014MNRAS.441..404H}, just
like ram pressure causes gas to lag a long way behind stars in the Bullet
Cluster \citep{2004ApJ...604..596C,2006ApJ...648L.109C,2008ApJ...679.1173R,2015Sci...347.1462H}.
Simple simulations tailored to Abell\,3827 support this prediction \citep[][although
current results operate under the limited assumption that the galaxy is on first
infall]{2015arXiv150406576K}.  Many particle physics models of dark matter naturally
predict low level
self-interactions \citep[e.g.][]{resonant,mirror,glueballs,SIMP,composite,2014JHEP...10..061K}.
If the interaction cross-section is considerable $\gs0.1$\,cm$^2$/g, it could also
resolve small-scale issues in the predictions of inert, cold dark matter (CDM)
models \citep[see review by][]{Weinberg02022015}.

Cluster Abell~3827 was originally studied by \cite{2011MNRAS.415..448W} because its
light distribution is interesting, with the intention of developing a lens analysis
algorithm but not with the expectation of measuring an offset (L.\ Williams 2015,
{\em pers.\ comm.}).  This is the only galaxy for which an offset has been detected,
but it may also be the only galaxy in a cluster for which such a small offset {\em
could} have been detected.  The measurement requires three chance circumstances, each
individually rare.

\begin{itemize}[topsep=0pt, itemindent=0.5em]

\item The cluster must gravitationally lens a well-aligned background galaxy with a complex morphology.
The distribution of foreground dark matter (plus baryons) can be reconstructed from
perturbations to this lensed image.

\item The cluster must contain a bright galaxy near the Einstein radius.
To enable precise measurements, it must intersect the lensed arcs and its mass must
be a detectable fraction of the cluster. Since a single cD galaxy generally lies
inside any Einstein rings, in practice, this means a cluster with multiple cDs.

\item The cluster must be nearby, so small physical separations can be resolved. This reduces its efficiency as a gravitational lens.

\end{itemize}

The interpretation of the observed offset in such radical terms as SIDM is clouded by
the possibility of alternative explanations.  First, gravitational lensing is
sensitive to the total mass distribution projected along the line of sight.  The
chance alignment of unrelated foreground/background structures has created apparently
spurious features in other lens
systems \citep{2001MNRAS.325..111G,2003MNRAS.339.1155H, 2012MNRAS.420L..18H}.
Similarly, source-lens degeneracies could lead to a similar
effect \citep[e.g.][]{Schneider2013, Schneider2014}.  In Abell~3827, projection
effects do not appear to be an issue: of the four galaxies at the centre of the
cluster three have a total mass appropriate for their stellar mass, while the fourth
(galaxy N1) has a low mass at the location of the stars, but a similarly appropriate
total mass slightly offset.  Had this been a chance projection, there would be mass
at the location of N1 (because of its own, non-offset dark matter) plus a second mass
peak (and probably a luminous source).  These are not seen.

Second, a physical offset might arise even with collisionless dark matter, via the
complex astrophysical processes operating in cluster core environments.  Gas stripped
from and trailing behind an infalling galaxy may self-gravitate and form new stars.
This is not consistent with observations of Abell~3827, which has effectively zero
star formation rate \citep[][Table~1]{Massey2015}.  The different physical extent of
dark matter and stars also leads to different dynamical friction, tidal gravitational
forces, and relaxation times during mergers.  Inside the complex distribution of
Abell~3827, even normally linear effects like tidal forces could create or exacerbate
small initial offsets.  It could also be considered that the galaxy in question is
undergoing one of two types of merger:
\begin{itemize}[topsep=0pt, itemindent=0.5em]

\item Coincidentally with the galaxy's arrival near the cluster core
, it has recently merged with a former satellite.  The tightly-bound stars from the
centre of the satellite have not yet had time to mix with the galaxy's stars, and
remain as a second peak randomly located within the total system.  Simulated
analogues of this are not consistent with observations, because the observed galaxy
is best-fit in all bands by a {\em single} Sersic
profile \citep[][Table~1]{Massey2015}.

\item The galaxy is about to merge with a more massive halo (the three more central galaxies of similar mass).
In simulations, the dark matter from all the systems rapidly mix together into a
single smooth halo.  This is not consistent with observations, which still show the
infalling galaxy's dark matter, {\em distinct from} and {\em further away from} the
other galaxies' dark matter\footnote{Allowing a distinct dark matter peak for N1 fits
the observations with $\chi^2\!/\mathrm{dof}\!=\!49.3/23$, Bayesian evidence
$\log_{10}{\!(E)}\!=\!-26.4$, and lens-plane
$\langle\mathrm{rms}_i\rangle\!=\!0.26\arcsec$ \citep[][Table~3]{Massey2015}. A model
without dark matter (but still stellar mass) is strongly disfavoured, with
$\chi^2\!/\mathrm{dof}\!=\!86.1/26$, $\log_{10}{\!(E)}\!=\!-100.7$ and
$\langle\mathrm{rms}_i\rangle\!=\!0.34\arcsec$ (R.\ Massey 2015, {\em
pers.\,comm}.).}.

\end{itemize}

As a control test to determine whether more complex astrophysical effects could build
an offset between galaxies and collisionless dark matter, we measure the 3D
separation between galaxies' luminous and dark matter in the ``Evolution and Assembly
of GaLaxies and their Environments'' (\eagle) suite of hydrodynamical cosmological
simulations \citep{Schaye2015, Crain2015}.  These simulations have been calibrated to
match the masses and sizes of galaxies in the local Universe.  The main \eagle
simulation also reproduces the observed low-redshift luminosity
functions \citep{Trayford2015} and produces an evolution of the galaxy mass function
in broad agreement with observations \citep{Furlong2015}.  The simulated galaxies
display rotation curves in agreement with observations whilst the stellar and dark
matter profiles of BCGs match observational
data \citep{Schaller2015a,Schaller2015b}. Similarly, the tidal stripping and ram
pressure stripping of the satellites as well as the AGN activity in the BCGs lead to
a realistic population of galaxies in clusters (in terms of colours or SFR),
indicating that the processes that could move matter around are broadly reproduced by
the model.  The \eagle simulations are hence, an ideal test-bed to predict the
relative positions of galaxies' various components in a statistically meaningful way.

\section{Method}
\label{sec:simulations}
In this section we describe briefly the cosmological simulations we analysed and the
method used to infer the centre of luminous and dark matter in galaxies.

\subsection{The simulation suite}

In our study, we use the main \eagle simulation (Ref-L100N1504) and to explore field
galaxies, clusters and groups, and the higher resolution simulation (Recal-L025N0752)
to understand the convergence of our results. These cosmological simulations use a
state-of-the-art treatment of smoothed particle hydrodynamics and set of subgrid
models. The full description of the model is given in \cite{Schaye2015} and the
rationale for its parametrisation is presented in \cite{Crain2015}; we only summarise
here the aspects relevant to our study. The simulations assume collisionless dark
matter, evolving in a flat \lcdm cosmology with parameters from \emph{Planck2013}
\citep{Planck2013}.  The low (high) resolution initial conditions are generated at
$z=127$ using second-order Lagrangian perturbation theory in a $100^3~\rm{Mpc}^3$
($25^3~\rm{Mpc}^3$) volume with a dark matter particle mass of $9.7\times10^6~\msun$
($1.2\times10^6~\msun$) and initial gas particle mass of $1.8\times10^6~\msun$
($2.2\times10^5~\msun$). The particles are then evolved in time using the
\textsc{Gadget} Tree-SPH code \citep{Springel2005}. The Plummer-equivalent
gravitational softening is set to $\epsilon=700~\rm{pc}$ ($\epsilon=350~\rm{pc}$ at
higher resolution).

The subgrid model in the \eagle simulations includes element-by-element radiative
cooling \citep{Wiersma2009a}, star formation obeying the Kennicutt-Schmidt relation
\citep{Schaye2008}, enrichment of the ISM via stellar mass loss \citep{Wiersma2009b},
feedback from star formation \citep{DallaVecchia2012}, gas accretion onto
super-massive black holes and the resulting AGN feedback
\citep{Booth2009,RosasGuevara2013}.

\subsection{Identification of galaxies and their locations}

We find galaxies in the simulation via the \textsc{Subfind} algorithm
\citep{Springel2001}.  We identify all galaxies with stellar mass
$M_*\!>\!10^9~\msun$ at $z\!=\!0$, both in the field and in groups or clusters.

We find the centre of galaxies' matter distributions using an iterative `shrinking
sphere'.  We first identify all the star particles for each galaxy. We calculate
their centre of mass and the distance of every particle to this centre. We then
select only those particles within $90\%$ of the maximal distance to the centre of
mass.  Repeating this process, the search radius and the number of considered
particles decreases in subsequent iterations.  This shrinking sphere procedure is
repeated until the number of particles reaches $200$. This typically corresponds to a
sphere of radius $\sim1~\rm{kpc}$, i.e. slightly larger than the softening length of
the simulation.  The centre of mass of this final set of particles is considered to
be the centre of the galaxy's stellar distribution\footnote{As pointed out by
  \cite{2014MNRAS.437.2865K,2015arXiv150406576K}, the choice of centroiding algorithm
  could produce varying results if dark matter does interact. Our identification of
  mass-weighted peaks in the stellar particles is both robust and the most comparable
  procedure to the identification of peaks in $K$-band luminosity-weighted
  observations (or other infrared bands in the absence of recent star formation).  }.
Similarly, we define the velocity of the stellar distribution as the mass weighted
velocity of the particles selected in the final iteration of the procedure.

The same procedure is applied to each galaxy's dark matter particles, to calculate
the centre of their dark matter distribution. Finally, the offset between the dark
and luminous component is defined as the distance between those two centres.  We have
verified that varying the minimum number of particles to define a galaxy centre from
100-500 and the shrinking ratio from 0.5-0.99 does not significantly affect our
results.


\section{Offsets between dark matter and stars} 
\label{sec:offset}
In the \eagle Ref-L100N1504 simulation there are $12776$ galaxies with mass
$M_*>10^9~\msun$, $1129$ of which are satellites in clusters (halos with
$M_{200}>10^{14}~\msun$), $3111$ satellites in groups (halos with
$M_{200}>10^{13}~\msun$) and $7391$ are field galaxies. The higher resolution
Recal-L025N0752 simulation contains $618$ galaxies above our mass threshold. These
four samples will be used to investigate environmental and resolution effects.

\subsection{3D offset between dark matter and stellar components}

\begin{figure}
\includegraphics[width=1.\columnwidth]{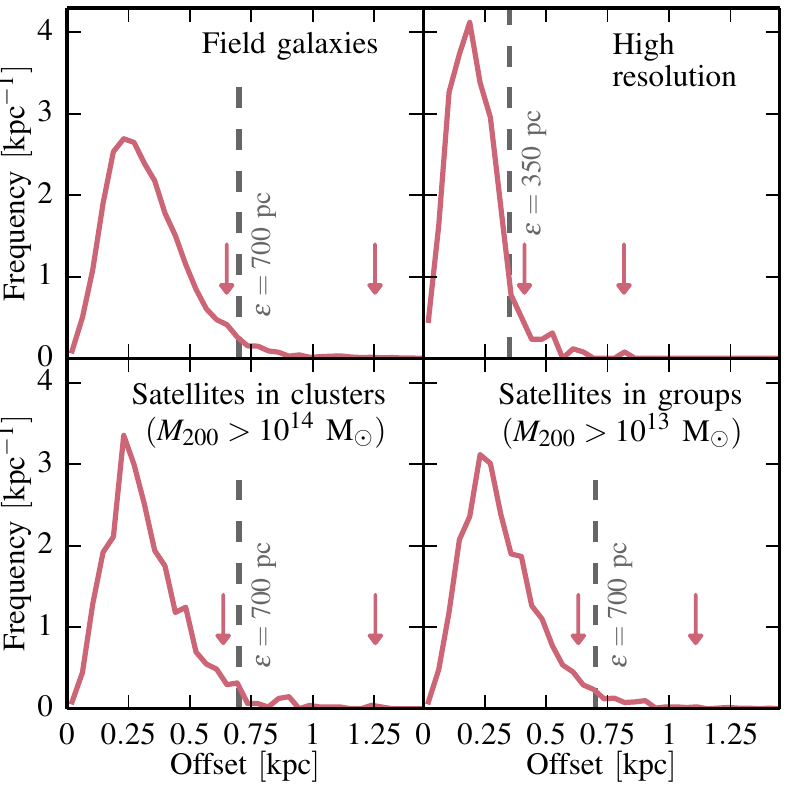}
\caption{The offset between the centre of the dark matter distribution and 
stellar distribution for galaxies with a stellar mass $M_*>10^9\msun$. The different
panels correspond to field galaxies in the reference simulation (top left), the field
galaxies in the higher resolution simulation (top right), satellites in clusters
(bottom left) and in groups (bottom right). In each panel the vertical dashed line
indicates the softening length used in the simulation. The arrows indicate the
position of the $95\%$ and $99.7\%$ percentiles of each distribution. The offset seen
is similar in field galaxies and clusters and is of the order of the softening scale
of the simulation. Offsets larger than $1.5~\rm{kpc}$ correspond to fluctuations
greater than $3\sigma$.}
\label{fig:offset}
\end{figure}

The offsets between the centre of galaxies' dark matter and their stars for our four
sub-samples of galaxies is shown in Fig.~\ref{fig:offset}.  The distributions are
consistent with being randomly sampled from a Maxwellian with distribution parameter
$\sigma\!=\!196\pm2$\,pc (main simulation) or $\sigma\!=\!126\pm1$\,pc (high
resolution simulation).  Arrows indicate the position of the $95$ and $99.7$ (2 and
3$\sigma$) percentiles.  In both cases, the typical scatter is smaller than the
gravitational softening length, indicated by a vertical dashed line.

The distribution of offsets in the Ref-L100N1504 main simulation is remarkably
similar for field galaxies (top left panel) and satellite galaxies in groups or
clusters (bottom panels).  This indicates that, at our resolution, the offsets are
not influenced by environmental effects. Fewer than $5\%$ of all galaxies display an
offset larger than the gravitational softening length.  Offsets larger than
$1~\rm{kpc}$ are only found in $59$ field galaxies ($0.79\%$) and 17 satellites in
groups and clusters ($0.54\%$). Pushing these numbers to offsets larger than
$1.5~\rm{kpc}$, we find $15$ field galaxies ($0.20\%$) and $2$ satellites in groups
and clusters ($0.06\%$). A much larger sample of galaxies would, however, be required
to characterise robustly the tail of the distribution.

Offsets in the higher resolution Recal-L025N0752 simulation (top right panel) are
smaller, with $95\%$ of the galaxies displaying an offset smaller than
$410~\rm{pc}$. Unfortunately, the smaller number of galaxies in that simulation
volume does not allow for a thorough discussion of the position of larger
percentiles. The results from this simulation indicate that the offsets seen in the
main simulation are probably overestimated (at least for field galaxies) and that
simulations run at a higher resolution (i.e.\ with a smaller softening length) would
lead to galaxies with smaller offsets between dark matter and stars. However, the
decrease in softening length by a factor of $2$ between our two simulations has only
led to a decrease in median offset by a factor of $1.5$, indicating that even higher
resolution simulations might not converge towards a negligible offset between
components\footnote{The offset of $300$-$400~\rm{pc}$ found by \cite{Kuhlen2013} in
high resolution zoom-in simulation of a single Milky Way-like galaxy is consistent
with our findings. That offset from the centre of their dark matter distribution is
$\sim\!3$ times larger than their softening, indicating that a small but non-zero
offset might be found with sufficiently high resolution adopted in the
simulations. We note, however, that the dark matter density profile of their galaxy
is not monotonic; a result different from what is seen in other simulations.}.  We
nevertheless caution that the softening length is not the only scale setting the
resolution of a simulation. Changes in the subgrid parameters and, sometimes, models
between different simulations at different resolution are necessary to account for
the newly resolved scales and have a non-trivial impact on the analysis of
convergence.

\subsection{Offset along the direction of motion}

If the dark matter-stellar offset in Abell~3827 is due to SIDM, then
not only will the centres of the galaxies and dark matter halos be
offset but this offset should also be aligned with the direction of
motion of the galaxy with the dark matter trailing the stars. Although
the offsets observed in the \eagle simulation are approaching the
resolution limit set by the scale of gravitational softening, it is
worth measuring whether the dark matter might be trailing or leading
the galaxies in their motion.

\begin{figure}
\includegraphics[width=1.\columnwidth]{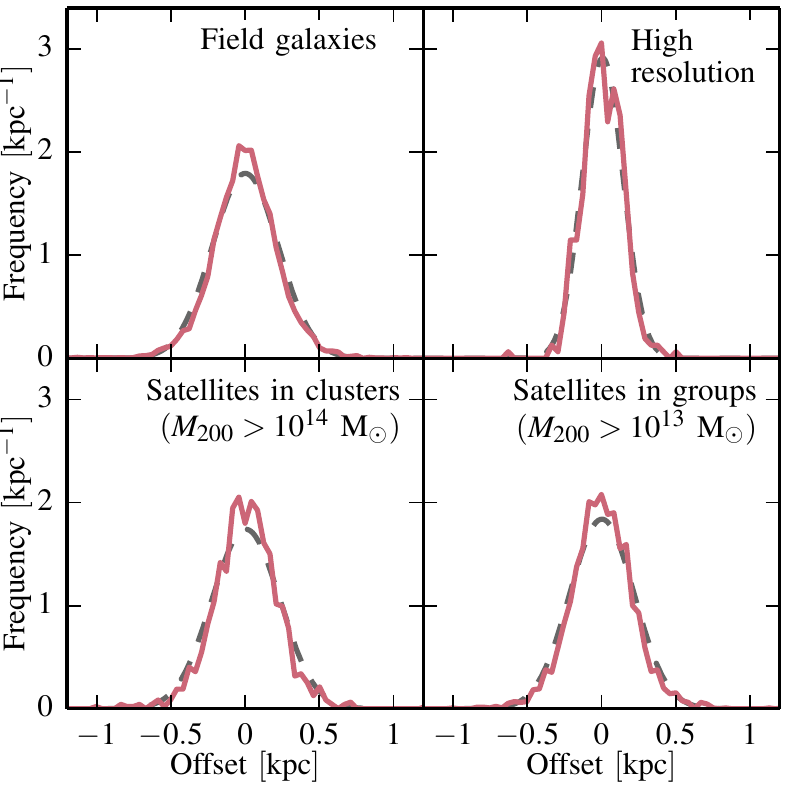}
\caption{The offset between the centre of the dark matter distribution and 
stellar distribution along the axis of motion of the stellar distribution for
galaxies with a stellar mass $M_*>10^9\msun$. The different panels correspond to the
same subsets of galaxies as in Fig.~\ref{fig:offset}. The dashed grey curves in the
background show a Gaussian distribution with the same mean and standard deviation as
the offset distributions.  The distribution of offsets displays no bias towards
trailing or leading motion of the dark matter centre with respect to the luminous
centre and only deviates from a normal distribution by displaying a positive
kurtosis. }
\label{fig:orientated_offset}
\end{figure}

The offset between dark matter and stars, projected along the velocity vector of the
stars, is shown in Fig.~\ref{fig:orientated_offset}.  In all four galaxy sub-samples,
the distribution is symmetric and shows no bias towards leading or trailing motion of
the dark matter. The distribution and its mirror image are indistinguishable in a
Kolmogorov-Smirnov (KS) test, with a $p$-value larger than $0.9$. The scatters are
$\sigma\approx210$\,pc (main simulation) or $\sigma\approx128$\,pc (high resolution
simulation), in agreement with those for the Maxwellian 3D offsets. The dashed lines
in the figure show Gaussians with the same mean and width as the measured
distributions. The distributions of projected offsets seem leptokurtic, with more
offsets near $0$ than the equivalent Gaussian, but the small number of galaxies in
the simulation does not probe the tails of the distribution. The offsets are, thus,
consistent with being randomly orientated and unaffected by the motion of the galaxy.
Indeed, we find no preferred direction of offset, repeating the experiment by
projecting the offset onto other axes like the dark matter velocity, direction to
nearest neighbour, direction to cluster centre, etc. All offsets are consistent with
random scatter.

\subsection{Detailed examination of satellite galaxies in the tail of the distribution}

In our sample of satellite galaxies, we found $17$ ($2$) objects out of $3111$
presenting an offset larger than $1~\rm{kpc}$ ($1.5~\rm{kpc}$). It is hence worth
exploring whether these are just random fluctuations in the population or whether
these larger offsets are seen as the result of an astrophysical process.  The offsets
of the 17 galaxies display no preferred direction with respect to the direction of
motion, with a flat distribution of $\cos{(\theta)}$, where $\theta$ is the angle of
the offset from the velocity vector of the galaxy.

The first of the two extreme outliers is a low mass ($M_*=3.9\times10^9~\msun$)
extended galaxy ($r_{50}=6.9~\rm{kpc}$). This galaxy is too diffuse at the resolution
of the simulation for the centre-finding algorithm to return a sensible
answer. Similarly, it would be difficult to find the centre of the light distribution
of a galaxy with such a flat profile in real observations. A galaxy like the one for
which an offset has been observed in Abell~3827 is much more massive and less
diffuse, making the presence of this specific outlier in our catalogue irrelevant for
the scenario we are considering.

The second extreme outlier is a giant elliptical galaxy with stellar mass
$M_*=1.5\times10^{11}~\msun$, located $130~\rm{kpc}$ from the centre of its
cluster. This galaxy experienced a recent merger with a smaller very concentrated
satellite ($M_*\approx2\times10^9~\msun$).

The dark matter from the two galaxies has mixed, forming a smooth, virialised halo.
The stars from the elliptical lie at the centre (within $200~\rm{pc}$) of this dark
matter.  However, the tightly-bound stars from the centre of the former satellite
have not yet had time to mix with the stars from the elliptical. They instead remain
as a peak in the outskirts of the stellar light distribution.  This merger remnant is
thus affecting the measurement of the peak of the light distribution but, at the time
of measurement, it does not carry any dark matter.  The large perceived offset is a
temporary phenomenon due to the difference between the time taken to mix the stars in
interacting galaxies and the time needed to mix their dark matter.  This is the first
merger scenario discussed in the Introduction, and is not consistent with details of
the observations (except perhaps a short time window might exist during which
distinct dark matter peaks still exist, but are offset from the light. This time
window would make Abell~3827 even more rare.)

\subsection{Detailed examination of satellite galaxies in the cores of clusters}

The simulation contains 50 (11) $M_*>\times10^{10}~\msun$ satellite galaxies within
the central $100$\,kpc of groups (clusters).  The statistics for this small sample are
noisier, but they have a similar offset distribution and distribution of angles
between offset and velocity vector as the full sample.  The distribution of angles is
consistent with uniform and the distribution of offsets has a mean of $310$\,pc with
a $95$~percentile at $690$\,pc, in remarkable agreement with the whole population.
This sub-sample and the whole population are virtually indistinguishable in a KS test
($p$-value $>\!0.6$).

The closest non-BCG galaxy with $M_*\gtrsim 10^{10}\msun$ in the six simulated
$M_{200}>10^{14}~\msun$ clusters are at clustercentric radii $26$\,kpc, $92$\,kpc,
$22$\,kpc, $58$\,kpc, $82$\,kpc and $54$\,kpc. These have offsets between their stars
and dark matter of $182$\,pc, $223$\,pc, $252$\,pc, $198$\,pc, $320$\,pc and
$284$\,pc, in apparently random directions. Looking in more detail at the two objects
with the smallest clustercentric position, we find two elliptical galaxies of mass
$1.5\times10^{10}$ and $4\times10^{10}\msun$ with low star formation rate and gas
content. They both present an offset between their dark and luminous component
smaller than $250$\,pc, unaligned with their direction of motion nor aligned with the
radius to the centre of the cluster. We thus find no feasible analogues for
Abell~3827 in the \eagle simulation.

\section{Summary \& Conclusion}
\label{sec:summary}
Motivated by the measurement of a $1.62^{+0.47}_{-0.49}$\,kpc offset between the
stars and dark matter of a galaxy in Abell 3827 \citep{Massey2015}, we
investigated the relative location of these matter components in galaxies from the
\lcdm \eagle simulation suite. Our results can be summarised as
follows:

\begin{enumerate}[labelwidth=3em, itemindent=2em, parsep=2mm]

\item 
More than $95\%$ of simulated galaxies have an offset between their
stars and dark matter that is smaller than the simulation's
gravitational softening length ($\epsilon=700~\rm{pc}$).  The offsets
are smaller still in our higher resolution simulation, indicating that
our measured values are likely upper limits.  Even this
state-of-the-art cosmological simulation has only just sufficient
resolution to compare to the observations.

\item Of the extreme objects with resolved offsets, fewer than
  $0.54\%$ ($0.20\%$) of satellite galaxies in groups and clusters
  present a separation larger than $1$\,kpc ($1.5$\,kpc).
 
\item We find no systematic alignment between the direction of the
  offset and the direction of motion of the galaxies. Dark matter is
  statistically neither trailing nor leading the stars.

\item We find no difference between field galaxies and satellite
  galaxies in groups and clusters. Astrophysical effects related to a
  galaxy's local environment play no significant role in producing or
  enhancing offsets.

\item We find two types of outliers with extreme offsets: faint
  galaxies for which the resolution of the simulations does not allow
  for the robust identification of a centre, and massive galaxies that
  have recently absorbed a smaller galaxy but haven't yet mixed their
  stellar distributions. Neither of these outlier types match what is
  observed in Abell 3827.

\item Looking specifically at the massive satellite galaxies close to
  cluster cores, we find no difference between these objects and the
  overall population of satellites or field galaxies. Environmental
  effects do not seem to create offsets.

\end{enumerate}

Astrophysical effects, as modelled in the \eagle simulation, produce
no feasible analogue, for the galaxy observed in Abell~3827.  Taking
the best-fitting value for its observed offset, this galaxy would be a
$>\!3\sigma$ outlier in a \lcdm universe with collisionless dark
matter.  Larger, higher-resolution simulations will, however, be
needed to improve the sampling of the tail of the offset distribution
and and to assess if the offsets measured in our simulation are biased
high by limited numerical resolution.

The observation is so far unique, and finding more systems in which
similarly precise measurements can be obtained will be challenging.
If more large offsets can be found and larger simulations confirm our
findings, the case for an alternative dark matter model (e.g.\ SIDM)
would be compelling.  High resolution simulations including these
models of dark matter would also be useful, to understand the
processes that might have led to the observed offset in Abell~3827.

\section*{Acknowledgements}
This work would have not be possible without Lydia Heck and Peter Draper's technical
support and expertise.  We also thank Doug Clowe, Alastair Edge, David Harvey,
Mathilde Jauzac, Tom Kitching, Prasenjot Saha, Liliya Williams and Idit Zehavi for
discussions about Abell~3827 and the \eagle team for giving us access to the
simulations.  RM is supported by the the Royal Society and the Leverhulme trust
(grant number PLP-2011-003).  This work was supported by the Science and Technology
Facilities Council (grant number ST/F001166/1); European Research Council (grant
numbers GA 267291 ``Cosmiway'').  This work used the DiRAC Data Centric system at
Durham University, operated by the Institute for Computational Cosmology on behalf of
the STFC DiRAC HPC Facility (www.dirac.ac.uk). This equipment was funded by BIS
National E-infrastructure capital grant ST/K00042X/1, STFC capital grant
ST/H008519/1, and STFC DiRAC Operations grant ST/K003267/1 and Durham
University. DiRAC is part of the National E-Infrastructure.  We acknowledge PRACE for
awarding us access to the Curie machine based in France at TGCC, CEA,
Bruy\`eres-le-Ch\^atel.

\bibliographystyle{mnras} 
\bibliography{./bibliography.bib}

\label{lastpage}

\end{document}